# WORST CASE BUFFER REQUIREMENTS FOR TCP OVER ABR [a]

B. Vandalore, S. Kalyanaraman[b], R. Jain, R. Goyal, S. Fahmy
*Dept. of Computer and Information Science,*
*The Ohio State University,*
*2015 Neil Ave,*
*Columbus,*
*OH 43210-1277, USA*
*E-mail: {vandalor, shivkuma, jain}@cis.ohio-state.edu*

ATM (asynchronous transfer mode) is the technology chosen for the Broadband Integrated Services Digital Network (B-ISDN). The ATM ABR (available bit rate) service can be used to transport "best-effort" traffic. In this paper, we extend our earlier work on the buffer requirements problem for TCP over ABR. Here, a worst case scenario is generated such that TCP sources send a burst of data at the time when the sources have large congestion windows and the ACRs (allowed cell rates) for ABR are high. We find that ABR using the ERICA+ switch algorithm can control the maximum queue lengths (hence the buffer requirements) even for the worst case. We present analytical arguments for the expected queue length and simulation results for different number of sources values and parameter values.

## 1 Introduction

ATM is designed to handle different kinds of traffic (voice, audio, video and data) in an integrated manner. ATM uses small fixed-size (53 bytes) packet (also called cells). It provides multiple service categories to support various quality of service (QoS) requirements. The current set of categories specified are: the constant bit rate (CBR), real-time variable bit rate (rt-VBR), non-real time variable bit rate (nrt-VBR), available bit rate (ABR), and unspecified bit rate (UBR). The CBR service is aimed at transporting voice and synchronous applications. The VBR (rt- and nrt-) services provide support for video and audio applications which do not require isochronous transfer. The ABR and UBR provide "best-effort" delivery for data applications. The ABR service uses closed-loop feedback to control the rate of sources.

The ATM technology is already being used in the backbone of the Internet. The performance of Internet protocols such as TCP/IP over ATM has been addressed in references [1,2]. ATM switches need buffers to store the packets before forwarding them to the next switch. We have addressed the problem of buffer requirements in the context of transporting TCP/IP applications over ABR [3,4,5]. In earlier studies, we had shown that achieving zero loss ABR service requires switch buffering which is only a small multiple of round trip times and the feedback delay. The buffering depends upon the switch scheme used.

---

[a] In Proc. of SICON'98, Singapore, June 1998
[b] S. Kalyanaraman is now with Dept. of ECSE, Rensselaer Polytechnic Institute, Troy, NY 12180-3590

One of the issues related to buffer sizing is that it is possible for a source to reach a high ACR and retain it. As long as it sends a packet before 500 ms elapse, the "use-it or lose-it" policy (the source loses its allocation if it does not use it to send data [6]) will not be triggered. The source can then use the high ACR to send a large amount of data suddenly. The effect can be amplified when many sources do the same thing.

In this paper, we generate a worst case scenario in which the TCP sources have a large congestion window and all ACR rates are high. Under such a condition, the TCP sources are made to send a huge burst of data into the network. We show that even under these extreme circumstances, ABR, using a good switch algorithm like ERICA+ [7,8], performs well and controls the queues. We present results of simulation of up to 200 TCP sources.

## 2  Congestion Control

In this section, we give a brief introduction to the TCP/IP congestion control and ABR flow control mechanisms.

### 2.1  TCP Congestion Control

TCP provides reliable, connection-oriented service. TCP connections provide window based end-to-end flow control [9]. The receiver's window (*rcvwnd*) is enforced by the receiver as a measure of its buffering capacity. The congestion window (*cwnd*) is used at the sender as a measure of the capacity of the network. The sender cannot send more than the minimum of *rcvwnd* and *cwnd*. The TCP congestion control scheme consists of the "slow start" and "congestion avoidance" phases. In the "slow start" phase, *cwnd* is initialized to one TCP segment. The *cwnd* is incremented by one segment for each acknowledgement received, so the *cwnd* doubles every round trip. The "congestion phase" is entered when *cwnd* reachs *ssthresh* (initially 64K bytes). In this phase the *cwnd* is incremented by 1/*cwnd* for every segment acknowledged. If an acknowledgement is not received by the time out period, the segment is considered to be lost, and "slow start" phase is entered. The *cwnd* is set to one, and *ssthresh* is set to *max(2, min(cwnd/2, rcvwnd))*.

## 2.2 ABR Flow Control

The ABR service uses closed-loop feedback control to advise the sources about the rate at which they should be transmitting the data. The switches monitor their load, compute the available bandwidth and divide it fairly among the competing connections. The feedback from the switches to the sources is sent in Resource Management (RM) cells which are sent periodically by the sources and turned around by the destinations (see Figure 1).

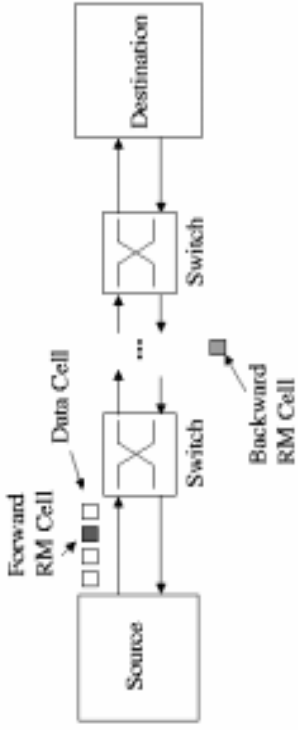

Figure 1: RM cell path

The RM cells flowing in the forward direction are called forward RM cells (FRMs) while those returning from the destination to the source are called backward RM cells (BRMs). When a switch receives a BRM cell, it computes its allowed cell rate (ACR) and sets it in the ER (explicit rate) field of that cell.

## 3 Generating Worst Case TCP Traffic

As previously mentioned, the maximum burst which TCP can send is equal to the maximum congestion window size. The TCP takes log($cwnd$) number of round trips to reach the maximum congestion window. In normal TCP over ABR activity when this maximum congestion window is achieved, the ACRs may not be high. In order to generate worst case conditions in which TCP has maximum congestion window size and the ACR is high, we have designed the following scenario.

A configuration where $N$ TCP sources connected to $N$ TCP destinations via two switches and a bottleneck link is used (see Figure 2).

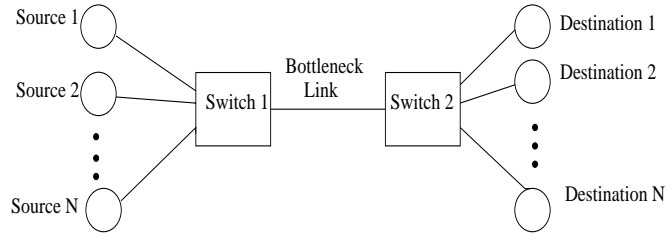

Figure 2: N Sources - N Destinations Configuration

Initially to build up the congestion window, each source sends one segment of data every $t$ seconds. One thousand such segments are sent by each source, so that congestion window reaches a maximum for all the sources. The sources send segments in a staggered manner, i.e., not all sources send the segments simultaneously. This is done so that the TCP data can be sent without overloading the network. Since the network is not overloaded, the ACRs are high and the congestion window reaches the maximum values. At this point ($1000 \times t$ seconds), all the sources synchronize and send a burst of data (burst size equal to maximum congestion window). This is the worst case burst size which can arise due to the $N$ TCP sources.

## 4  TCP Options and ERICA+ parameters

We use a TCP maximum segment size (MSS) of 512 and 1024 bytes. The MTU size used by IP is generally 9180 bytes, so there is no segmentation caused by IP. Since our simulations are performed under no loss conditions, the results hold even for TCP with fast retransmit and recovery or TCP with SACK (Selective Acknowledgements).

The TCP data is encapsulated over ATM as follows. First, a set of headers and trailers are added to every TCP segment. We have 20 bytes of TCP header, 20 bytes of IP header, 8 bytes for the RFC1577 LLC/SNAP encapsulation, and 8 bytes of AAL5 information, for a total of 56 bytes. Hence, every MSS of 512 bytes becomes 568 bytes of payload for transmission over ATM. This payload with padding requires 12 ATM cells of 48 data bytes each.

The ERICA+ algorithm operates at each output port (or link) of a switch. The switch periodically monitors the load on each link and determines quantities such as, load factor, the ABR capacity, and the number of active virtual connections or VCs. A measurement or "averaging interval" is used for this purpose. These quantities are used to calculate the feedback which is indicated in BRM cells. The measurements are made in the forward direction and feedback is given in reverse direction. ERICA+ uses dynamic queue control to quickly drain queues if there is a large queue build up. A hyperbolic queue control function is used to vary the available ABR capacity as a function of the current queue length.

The ERICA+ algorithm uses the queuing delay as a metric to calculate the feedback. ERICA+ uses four parameters: a target queuing delay (T0 = 500 microseconds), two curve parameters (a = 1.15 and b = 1.05), and a factor which limits the amount of ABR capacity allocated to drain the queues (QDLF = 0.5). In our simulations, the exponential averaging options for averaging N (number of active VC's) and for averaging overload are used to smooth the variances in the traffic.

## 5 Analytical Explanation

In this section, we derive the analytical value for the maximum queue lengths as a function of the number of sources, congestion window size and congestion status of network. The queue length is given in two equation for underloaded and overloaded network conditions as follows:

$$Queue_{under} = N \lfloor \frac{cwnd\_max}{48} \rfloor \qquad \text{for} \quad N \leq \lfloor \frac{t}{g} \rfloor \qquad (1)$$

$$Queue_{over} = 353356 \times N \times t \qquad \text{for} \quad N > \lfloor \frac{t}{g} \rfloor \qquad (2)$$

where

- $N$ = number of sources
- $cwnd\_max$ = maximum congestion window size
- $t$ = time between consecutive segments
- $g$ = time between sources sending segments for all sources

(353356 is the number of cells sent in the burst by one source under overloaded condition. This is explained in the derivation of equation 2 later in this section)

The derivation of the two equations is as follows: Initially, for a small number of sources, the network is underloaded, so the ACRs are high when the burst occurs. When all the sources send the burst simultaneously, the whole burst can be sent into the network and this results in large switch queues.

Let $cells\_in\_mss$ be the number of cells required to send a segment and $mss$ be the maximum segment size.

The burst size due to one source = $\lfloor cells\_in\_mss \times \frac{cwnd\_max}{mss} \rfloor$ cells

So if there are N sources, the expected queue length is

$$\text{Queue length} = N \lfloor cells\_in\_mss \times \frac{cwnd\_max}{mss} \rfloor \text{cells}$$

Substituting $cells\_in\_mss = \lceil \frac{mss}{48} \rceil$ in the above we get (equation 1)

$$\text{Queue length} = N \lfloor \frac{cwnd\_max}{48} \rfloor \quad (1)$$

Under the no loss condition, the TCP congestion window increases exponentially in the "slow start" phase until it reaches the maximum value. As the number of sources increase the load of the network increases. In the simulation, every source sends a segment of size 512 bytes every $t$ seconds. The time between two sources sending their segments is $g$ seconds. At time 0, source 1 sends a segment, then at time $g$ seconds, source 2 sends, at time $2g$ seconds source 3 sends and so on. Also source 1 sends its second segment at time $t$ seconds and so on.

The network becomes overloaded when the input rate is greater than the output rate. In the simulation, the network will get overloaded if number of sources

$$N > \lfloor \frac{t}{g} \rfloor$$

for $t = 1$ millisecond, $g = 50$ microseconds, $N = \frac{1000 \times 1}{50} = 20$.

Once the network is overloaded, the ACRs become low and each of the N sources gets approximately $\frac{1}{N}$ of the bandwidth. Since the ACRs are low, the burst which occurs after 1 second should not give rise to large queues. There are still switch queues which occur initially when the network has not yet detected that it is overloaded.

The TCP congestion window grows exponentially whenever it receives an acknowledgement. Let $d$ be the length of the links in kms. The acknowledgement arrives after a round trip time of $30\frac{d}{1000}$ ms. But, by this time the source will have generated more segments to send. After each round trip, each source will send twice the number of segments. A larger round trip can give rise to a larger burst. The network gets overloaded when the number of segments sent just fills up the pipe.

Let $k$ be the number of round trips it takes to overload the network. Time taken for transmitting one cell is 2.83 microsecond (at 149.76 Mbps, on an OC-3 link accounting for SONET overhead). In $t$ seconds, each source sends one segment which gives rise to $cells\_in\_mss$ cells. Therefore, the network gets overloaded when

$$2^k \times cells\_in\_mss = \text{Number cells sent in t seconds}$$

$$\implies k = \lceil \lg \frac{t \times 10^6}{2.83 \times cells\_in\_mss} \rceil$$

For $t = 1$ millisecond, $mss = 512$, the value of $k$ is 5. Hence, in this case after 6 ($= k+1$) round trips, the network detects the overload if $N > 20$. The maximum queues should occur after 6 round trips for $N > 20$.

The value of the maximum queue in an overloaded network can be derived as follows.

The burst due to one source $= 2^k \times cells\_in\_mss = \frac{t \times 10^6}{2.83} = 353356 \times t$ cells.

Hence, the burst due to $N$ sources $= 353356 \times N \times t$ cells $\qquad$ (2)

For $t = 1$ millisecond, $g = 50$ microseconds, $mss = 512$ and $cwnd\_max = 65536$, we get

$$Queue_{under} = 1365 \times N \qquad \text{for} \quad N \leq 20 \qquad (3)$$

$$Queue_{over} = 353 \times N \qquad \text{for} \quad N > 20 \qquad (4)$$

## 6 Simulation Results

We show the maximum queues at the bottleneck switch for the cases where the number of sources $N$ varies from 2 to 200. The results are shown in Table 1. The plot of the expected queue length and the actual length obtained from simulation results versus the number of sources is shown in Figure 3.

The length of all the links is 1000 km, giving rise to 15 ms propagation delay from source to destination. The round trip time (RTT) is 30 milliseconds. The feedback delay (the delay between the bottleneck link and the source and back) is 10 ms. A round trip of 30 ms corresponds to 11029 cells. The other parameter values are $t = 1$ millisecond, $g = 50$ microseconds, $mss = 512$ bytes, $cwnd\_max = 65536$.

The maximum queue length for sources $N \leq 20$ was at time around 1 second. For $N > 20$, the maximum queue was at time around 300 millisecond (which agrees with the analytical prediction). The queue lengths increase with the number of sources as expected. The queue lengths also correspond to the burst size sent by the sources. But, this is true only when the number of sources is less than or equal to 20. For $N = 30$, there is a sharp decrease in the maximum queue length. For $N > 40$, it starts increasing again but now it increases at a slower rate.

From Table 1, it can be seen that for $N \leq 20$, the simulation agrees well with the analytical values. For $N > 30$ initially the queue lengths in the simulation are higher than the analytical value, but later it becomes lower than the expected queue length. This can be explained as follows. Initially the traffic generated in each cycle ($t$ seconds) can be sent within that cycle. But as the number of sources increases, ($N > 80$) the network gets overloaded, and there are some pending cells which get sent in the next cycles. Because of this the network detects the overload at lower queue lengths than the analytical value.

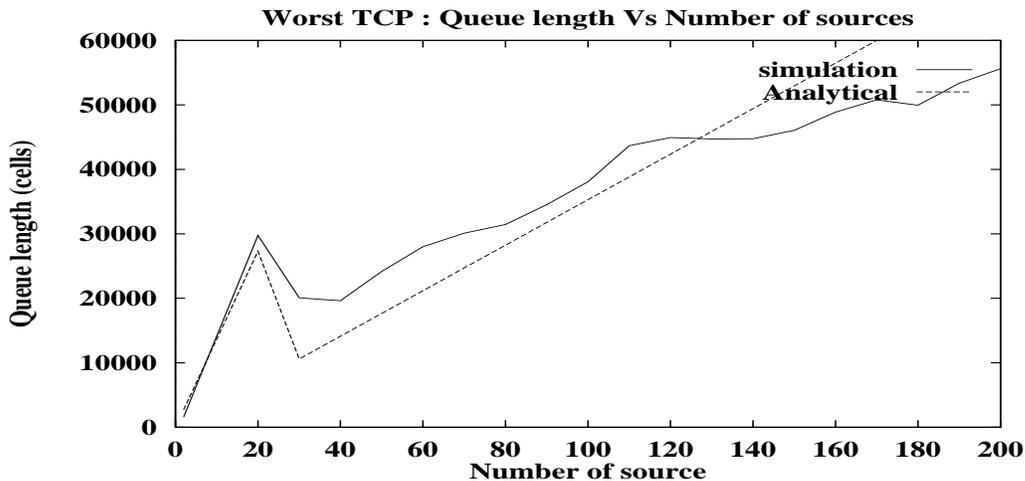

Figure 3: Queue length Versus Number of Sources

To study the effect of different parameters, a full factorial experiment was carried out for different parameter values. For a given number of sources the following parameters were varied: maximum segment size ($mss = 512, 1024$ bytes), time between two sources sending segments ($g = 50, 100$ microseconds), time between successive segments of a source ($t = 1, 10$ milliseconds) and length of the links ($d = 1000, 2000$ kms).

Table 2 shows for number of source $N = 3, 10, 30, 40, 50, 100$ the maximum queue sizes for 16 experiments. From Table 2, the following observations can be made. (These observations support the analytical results.)

- In line numbers 10 and 14 for $N = 3, 10$ the queue size is small compared to other lines, since in these cases congestion is detected early, so the ACRs are low when the burst is sent. Queue length is given by the equation 2.

- For $N = 30, 40, 50, 100$, when $t = 10$ millisecond, the queue lengths agrees with the one given by the equation 1.

Table 1: Effect of the number of sources: *actual* and *expected* queue lengths

| # TCP Sources | Max Queue size (cells) | Analytical Queue Size (cells) | # TCP Sources | Max Queue size (cells) | Analytical Queue Size (cells) |
|---|---|---|---|---|---|
| 2 | 1575 | 2730 | 100 | 38088 | 35300 |
| 3 | 3149 | 4095 | 110 | 43672 | 38830 |
| 5 | 6297 | 6825 | 120 | 44939 | 42360 |
| 10 | 14131 | 13650 | 130 | 44708 | 45890 |
| 20 | 29751 | 27300 | 140 | 44744 | 49420 |
| 30 | 20068 | 10590 | 150 | 46058 | 52950 |
| 40 | 19619 | 14120 | 160 | 48880 | 56480 |
| 50 | 24162 | 17650 | 170 | 50784 | 60010 |
| 60 | 28006 | 21180 | 180 | 49961 | 63540 |
| 70 | 30109 | 24710 | 190 | 53366 | 67070 |
| 80 | 31439 | 28240 | 200 | 55618 | 70600 |
| 90 | 34530 | 31770 | - | - | - |

Table 2: Effect of MSS ($mss$), $Distance$(d), time intervals ($t, g$)

| # | $mss/g/t/d$ | N=3 | N=10 | N=30 | N=40 | N=50 | N=100 |
|---|---|---|---|---|---|---|---|
| 1 | 512/50/1/1000 | 3171 | 14273 | 20068 | 19619 | 24162 | 35687 |
| 2 | 512/50/1/2000 | 3171 | 14273 | 19906 | 27567 | 30872 | 75083 |
| 3 | 512/50/10/1000 | 3172 | 14274 | 45994 | 61854 | 77714 | 150453 |
| 4 | 512/50/10/2000 | 3172 | 14274 | 45994 | 61854 | 77714 | 150458 |
| 5 | 512/100/1/1000 | 3171 | 14273 | 19283 | 20080 | 24164 | NA |
| 6 | 512/100/1/2000 | 3171 | 14273 | 21241 | 32314 | 35961 | NA |
| 7 | 512/100/10/1000 | 3172 | 14274 | 45994 | 61854 | 77714 | NA |
| 8 | 512/100/10/2000 | 3172 | 14274 | 45994 | 61854 | 77714 | NA |
| 9 | 1024/50/1/1000 | 3040 | 13680 | 18650 | 18824 | 23542 | NA |
| 10 | 1024/50/1/2000 | 1542 | 5612 | 19131 | 22934 | 29163 | NA |
| 11 | 1024/50/10/1000 | 3040 | 13680 | 44080 | 59280 | 74480 | NA |
| 12 | 1024/50/10/2000 | 3041 | 13681 | 44081 | 59281 | 74481 | NA |
| 13 | 1024/100/1/1000 | 3040 | 13680 | 18591 | 19600 | 24314 | NA |
| 14 | 1024/100/1/2000 | 1403 | 5556 | 17471 | 24412 | 30533 | NA |
| 15 | 1024/100/10/1000 | 3040 | 13680 | 44080 | 59280 | 74480 | NA |
| 16 | 1024/100/10/2000 | 3041 | 13681 | 44081 | 59281 | 74481 | NA |

(NA - data not available)

- If the network is overloaded, then a larger round trip time gives larger queue lengths. (e.g., see line 1 and 2 for $N = 30, 40, 50$).

- Lines 3 and 4 indicate that in underloaded conditions the queue length is given by the equation 1.

- As expected, the segment size of 512 and 1024 do not affect the queue sizes.

## 7 Conclusion

We have artificially generated a worst case scenario and studied the buffer requirements for TCP over ABR. The buffer size required depends on the switch scheme used. Our simulations are based on our ERICA+ switch scheme. For the worst case scenario generated, both the analytical prediction and simulation results for queue lengths (and hence buffer requirements) are given. The simulation agrees well with the analytical prediction. The queue lengths are *affected* by *maximum congestion window size, the round trip time, network congestion* (overloaded or underloaded) and *number of sources*. It is *not affected* by the *maximum segment size*.

### Acknowledgments

This research was sponsored in part by Rome Laboratory/C3BC Contract #F30602-96-C-0156.

---

[c]All our papers and ATM Forum contributions are available through http://www.cis.ohio-state.edu/~jain/